\documentclass[11pt]{article} 
\usepackage{fullpage}
\usepackage{amssymb}

\newtheorem{theorem}{Theorem}
\newtheorem{definition}[theorem]{Definition}

\newtheorem{corollary}[theorem]{Corollary}

\newenvironment{remark}{\begin{trivlist}%
\item[\hskip\labelsep{\bf Remark.}]~}{\end{trivlist}}
\newenvironment{proof}{\noindent{\it Proof.  }}{\hspace*{\fill}
  \rule{2mm}{2mm} \vspace{5mm}} 

\newenvironment{claim}{\noindent{\bf Claim.}\em}{\\[3mm]}

\newcommand{\CNF}{\mbox{\rm CNF}}
\newcommand{\no}{\mbox{\rm NO}}
\newcommand{\yes}{\mbox{\rm YES}}

\renewcommand{\O}{\tilde{O}}
\newcommand{\proj}[2]{#1|_{#2}}
\newcommand{\Z}{{\mathbb{Z}}}
\newcommand{\ov}[1]{\ensuremath{\overline{#1}}}

\newcommand {\ket} [1] {\ensuremath{\vert}{#1}\ensuremath{\rangle}}

\newcommand{\xy}{\proj{x}{y}}
\newcommand{\xybar}{\proj{x}{\overline{y}}}

\title{The Quantum Query Complexity of 0-1 Knapsack and Associated Claw 
Problems}

\author{\begin{tabular}{cc}
V.~Arvind & \quad Rainer Schuler\quad \\
Institute of Mathematical Sciences  & \quad Theoretische Informatik\quad\\
C.~I.~T. Campus & \quad Universit\"at Ulm\quad\\
Chennai 600 113, India & \quad D-89069 Ulm, Germany\quad\\
{\tt arvind@imsc.res.in} & {\tt rsc@informatik.uni-ulm.de}\quad
\end{tabular}}  

\begin{document}

\date{}

\maketitle

\begin{abstract} 
  We first give an $\O(2^{n/3})$ quantum algorithm for the 0-1
  Knapsack problem with $n$ variables. More generally, for 0-1 Integer
  Linear Programs with $n$ variables and $d$ inequalities we give an
  $\O(2^{n/3}n^d)$ quantum algorithm. For $d =o(n/\log n)$ this
  running time is bounded by $\O(2^{n(1/3+\epsilon)})$ for every
  $\epsilon>0$ and in particular it is better than the $\O(2^{n/2})$
  upper bound for general quantum search.
  
  To investigate whether better algorithms for these NP-hard problems
  are possible, we formulate a \emph{symmetric} claw problem
  corresponding to 0-1 Knapsack and study its quantum query
  complexity.  For the symmetric claw problem we establish a lower
  bound of $\O(2^{n/4})$ for its quantum query complexity. We have an
  $\O(2^{n/3})$ upper bound given by essentially the same quantum
  algorithm that works for Knapsack.
  
  Additionally, we consider CNF satisfiability of CNF formulas $F$
  with no restrictions on clause size, but with the number of clauses
  in $F$ bounded by $cn$ for a constant $c$, where $n$ is the number
  of variables. We give a $2^{(1-\alpha)n/2}$ quantum algorithm for
  satisfiability in this case, where $\alpha$ is a constant depending
  on $c$.


\end{abstract}

\section{Introduction}\label{intro}

The goal of the present paper is to explore the possibility of
developing quantum algorithms for different NP-hard problems that are
faster than what we get by a direct application of Grover's search
algorithm \cite{grov}. This can be seen as part of an active research
theme in quantum computing: discovering tasks for which there are
quantum algorithms that are significantly faster than classical
algorithms (c.f.~\cite{grov,buhretal}). The difficulty in the area is
that there are as yet only a few known techniques for designing
quantum algorithms.  The two central methods are essentially from
Shor's factoring algorithm \cite{shor} and Grover's search algorithm
\cite{grov}. Other known algorithms are essentially based on these
methods.

In this paper, we consider quantum algorithms for certain NP-hard
problems which have a special divide-and-conquer structure that can be
exploited to design significantly faster quantum algorithms than what
a direct application of Grover's search algorithm would yield.  For
example, we give an $\O(2^{n/3})$ quantum algorithm for the 0-1
Knapsack problem with $n$ variables. This is a consequence of a more
general result for 0-1 Integer Linear Programs with $n$ variables and
$d$ inequalities for which we give an $\O(2^{n/3}n^d)$ quantum
algorithm. For $d =o(n/\log n)$ this running time is bounded by
$\O(2^{n(1/3+\epsilon)})$ for every $\epsilon>0$ and in particular it
is significantly better than the $\O(2^{n/2})$ time bound if we
directly apply Grover's search algorithm. Our algorithms are, of
course, based on Grover's search algorithm and the powerful method of
amplitude amplification \cite{grov,bhmt}.
  
The next question we address is whether faster quantum algorithms for
the above NP-hard problems are possible. To study this, we formulate a
new claw-like problem (see \cite{buhretal} for claw problems) which we
term as the \emph{symmetric} claw problem and study its quantum query
complexity.  The symmetric claw problem essentially captures the
structure of 0-1 Knapsack and we study its complexity in the quantum
query model \cite{bbbv,ambain}. For the symmetric claw problem we are
able to prove a lower bound of $\O(2^{n/4})$ in the quantum query
model using Ambainis' method \cite{ambain}. The problem also has an
$\O(2^{n/3})$ upper bound given by essentially the same quantum
algorithm that works for 0-1 Knapsack.
  
Finally, we consider CNF satisfiability for the case of CNF formulas
$F$ with number of clauses linearly bounded in the number of
variables. Note that we do not make any assumptions on the size of
clauses in $F$. More precisely, let $\CNF_c$ denote CNF formulas $F$
such that the number of clauses in $F$ is bounded by $cn$ for the
given constant $c$, where $n$ is the number of variables. We give a
$2^{(1-\alpha)n/2}$ quantum algorithm for satisfiability of inputs
from $\CNF_c$, where $\alpha$ is a constant depending on $c$.

\subsection{Preliminaries}

The set $\{0,1\}^n$ denotes the set of binary strings of length $n$,
and for a positive integer $N$ we denote the set $\{1,2,\ldots,N\}$
by $[N]$.

We design new quantum algorithms for the following NP-hard problems.\\

\noindent{\bf 0-1 Knapsack}\\

Input: A list of positive integers $c_1,c_2,\ldots,c_n$ and a positive
integer $K$. 

Problem: Is there a subset $S\subseteq[n]$ such that $\sum_{i\in S}
c_i=K$.\\

Actually, we consider is 0-1 Integer Linear Programs in general.

\noindent{\bf 0-1 ILP}

Input: Integers $a_{ij}$ and $b_i$, $1\leq j\leq n$ and $1\leq i\leq
d$. 

Problem: Is the following set of linear inequalities feasible?
\[
\sum_{j=1}^n a_{ij}x_j \leq b_i,~~~ x_j\in\{0,1\},~~~1\leq j\leq n ~~~ 
1\leq i\leq d.
\]

We use standard definitions from quantum computing from
\cite{bbbv,clev}. In particular, we use the quantum query model (c.f.
\cite{buhretal,ambain}). In this model the input is a function
$f:[N]\rightarrow [M]$ and the values of $f$ is accessed by oracle
queries. The complexity of computing some property of $f$ is measured
as the number of oracle queries. A quantum computation with $T$
queries can be seen as a sequence of unitary transformations:
\[
U_0\rightarrow O\rightarrow U_1\rightarrow\cdots O\rightarrow U_T,
\]
where the unitary transform $O$ implements the oracle access to $f$
and the $U_i$ are arbitrary unitary transformations which do not
depend on the input $f$.

A central idea from quantum computing, which is a generalized form of
Grover's search algorithm that we use throughout the paper is
amplitude amplification \cite{bhmt}: Essentially, if we have a quantum
algorithm ${\cal A}$ with success probability $p$ then the success
probability can be amplified to a constant by $O(\sqrt{1/p})$ calls to
${\cal A}$ and ${\cal A}^{-1}$.

A \emph{range tree} (see \cite{shamos} for details) is a data
structure for storing a set $S$ of $N$ elements
$\{x_1,x_2,\ldots,x_N\}$, each of which is a $d$-tuple of integers
$x_i=(x_{i1},x_{i2},\ldots,x_{id})$. The data structure can be be
built in time $N(log^{d-1} N)$ and needs space $N(log^{d-1} N)$.  The
specific property of interest that we require is that we can process
\emph{range queries} on a range tree in time $\log^d N$ to retrieve
one element, if it exists, of the set $S$ that satisfies the query
bounds. More precisely, given a $d$-tuple of real numbers
$(a_1,a_2,\ldots,a_d)$, in time $O(\log^d N)$ we can search for an
element $x_i=(x_{i1},x_{i2},\ldots,x_{id})$ in the range tree such
that $x_{ij}\leq a_j$, $1\leq i\leq d$.

\section{New Quantum Algorithms for 0-1 Integer Linear Programs}\label{ilp}

\begin{theorem}\label{theorem1}
  There is an $\O(2^{n/3}n^d)$ quantum algorithm with constant error
  probability that solves 0-1 integer linear programs with $n$
  variables and $d$ inequalities.
\end{theorem}

\begin{proof}
Let the input instance be
\[
\sum_{j=1}^n a_{ij}x_j \leq b_i, ~~~~ 1\leq i\leq d,\\
x_j\in\{0,1\}, ~~~~ 1\leq j\leq n.
\]

The goal is to find a feasible solution. We give a stepwise
description of the quantum algorithm.

\begin{enumerate}
\item[Step 1.] Partition the variables into two sets
  $A=\{x_1,\ldots,x_{n/3}\}$ and $B=\{x_{n/3+1},\ldots,x_n\}$,
where note that $|A|=n/3$ and $|B|=2n/3$.

\item[Step 2.] For each of the $2^{n/3}$ 0-1 assignments $I$ to the
  variables in $A$ define the $d$-tuple $y_I=(y_I(1).\ldots,y_I(d))$
of integers, where
\[
y_I(i)=\sum_{j=1}^{n/3} a_{ij}I_j,
\]
where $I_j$ is the value of variable $x_j$ in assignment $I$.

\item[Step 3.] Let $N=2^{n/3}$. The set $X=\{y_I\mid I$ is a 0-1
  assignment to $A\}$ is of size $N$. In time $\O(N\log^{d-1} N)$
  build a range tree (c.f. \cite[Theorem 2.11]{shamos}) to store the
  set $X$. The range tree has size $\O(N\log^{d-1} N)$ and range
  queries can be processed in time $\log^d N$. This completes the
  preprocessing phase.

\item[Step 4.] Using the Hadamard transform prepare the uniform
  superposition over the set $T$ of 0-1 assignments to the $2n/3$
  variables in $B$:
\[
\ket{\psi}=\frac{1}{2^{n/3}}\sum_{a\in T} \ket{a}.
\]

\item[Step 5.] Define a unitary transform $U$ as the standard
  reversible implementation of the following classical subroutine $f$
  given $u\in T$ as input:
\begin{itemize}
\item[(i)] Compute $z_j = \sum_{j=n/3+1}^n a_{ij}u_j$, $1\leq i\leq
  d$.
\item[(ii)] Let $\hat{z_i}=b_i-z_i$ for $1\leq i\leq d$, giving a
$d$-tuple of integers $(\hat{z_1},\ldots,\hat{z_d})$.
\item[(iii)] Search in the range tree for a $y_I$ such that
  $y_I(1)\leq \hat{z_1},\ldots,y_I(d)\leq \hat{z_d}$.  Notice that
  this is a range query and it can be processed in time $\log^d N$ in
  the range tree date structure \cite[Theorem 2.11]{shamos}.
\item[(iv)] If such a tuple $y_I$ is in $X$ then $f(u)=1$.  Otherwise,
  $f(u)=0$.
\end{itemize}

\item[Step 6.] With the initial state as the uniform superposition
  $\ket{\psi}$ and using the unitary transform $U$, apply Grover's
  search algorithm to search for $u\in T$ such that $f(u)=1$.
\end{enumerate}

It follows from the well-known analysis of Grover's algorithm and
amplitude amplification \cite{bhmt,grov} that the above algorithm has
running time $\O(2^{n/3}\log^d N) =\O(2^{n/3}n^d)$, with constant
success probability.
\end{proof}

\begin{corollary}
  There is an $\O(2^{n/3})$ time quantum algorithms with constant
  success probability for 0-1 Knapsack. Additionally, any NP-hard
  optimization problem that takes the form of a 0-1 Integer Linear
  Program with constant number of constraints has an $\O(2^{n/3})$
  time quantum algorithms with constant success probability.
\end{corollary}

\begin{proof}
  As an instance of 0-1 Knapsack consists of one equation, it can be
  expressed as a 0-1 integer linear program with two inequalities.
  Hence the algorithm of Theorem~\ref{theorem1} yields the claimed
  quantum algorithm.
  
  Consider any optimization problem that has a linear optimality
  function and the constraints can be expressed as a 0-1 integer
  linear program with $d$ constraints. Using binary search we can
  reduce the optimization problem to feasibility of a 0-1 integer
  linear program with $d+1$ constraints, which can be solved using the
  algorithm of Theorem~\ref{theorem1} with running time $\O(2^{n/3})$
  for constant $d$. Thus, the corresponding optimization problem can
  also be solved in time $\O(2^{n/3})$ with constant success
  probability.
\end{proof}

\begin{remark}
Interestingly, if the 0-1 ILP is of the following form:

\begin{eqnarray*}
\mbox{minimize/maximize~~}~ & \sum_i c_ix_i & \\
~~\mbox{ s.t. } & \sum_{j=1}^n a_{ij}x_j = b_i, & ~ 
1\leq i\leq d, x_j\in\{0,1\},~~ 1\leq j\leq n,
\end{eqnarray*}

then it is easy to see that essentially the same quantum algorithm
presented in Theorem~\ref{theorem1} solves this optimization problem
in time $\O(2^{n/3})$, independent of the number of equations $d$. The
reason is that we do not have to maintain a range tree data structure
as we do not have to process range queries. Just a sorted list would
suffice as we only need to make exact queries in this case.

The above kind of 0-1 ILP is referred to as the 0-1 Group Problem in
Nemhauser and Wolsey's book \cite{NeWo}. An interesting instance of
this problem is the following NP-hard problem: given a CNF formula $F$
(no restrictions on clause size), search for a satisfying assignment
that satisfies \emph{exactly} one variable in each clause. This is
clearly an instance of the above 0-1 ILP. Thus we have the following
corollary.
\end{remark}

\begin{corollary}
  There is an $\O(2^{n/3})$ quantum algorithm that takes a CNF formula
  $F$ with $n$ variables as input (no restrictions on clause size or
  number of clauses) and searches, with constant success probability,
  for a satisfying assignment that satisfies exactly one variable in
  each clause.
\end{corollary}

\section{The Symmetric Claw Problem}\label{claw1}

In order to study how far we can exploit this idea of dividing the
input, we examine in this section a black-box version of the 0-1
Knapsack problem which we term as the Symmetric Claw problem.

For a pair of strings $x,y\in\{0,1\}^n$ let $\xy$ denote the
substring of $x$ obtained by projecting it on the positions where $y$
is 1. Likewise, let $\xybar$ denote the substring of $x$ obtained by
projecting it on the positions where $y$ is 0.

\begin{definition}\label{claw1def}
The input to the \emph{symmetric claw problem} is a function
$P:\{0,1\}^n\times\{0,1\}^n\rightarrow \{0,1\}^m\times\{0,1\}^m$ where
we write $P(x,y)=(P_1(x,y),P_2(x,y))$ for every $x,y\in\{0,1\}^n$.
Additionally, the input function $P$ is such that it fulfills the
following promise symmetry conditions:

\begin{enumerate}
\item For $x\in\{0,1\}^n$, if $P_1(x,y)=P_2(x,y)$ for some
$y\in\{0,1\}^n$ then $P_1(x,y)=P_2(x,y)$ for every $y\in\{0,1\}^n$.

\item For any $x, x', y'\in\{0,1\}^n$ if $\proj{x}{y}=\proj{x'}{y}$ then
$P_1(x,y)=P_1(x',y)$. Likewise, if
$\proj{x}{\overline{y}}=\proj{x'}{\overline{y}}$ then
$P_2(x,y)=P_2(x',y)$.
\end{enumerate}

Given such an input $P$ the symmetric claw problem is to find an
$x\in\{0,1\}^n$, if it exists, such that $P_1(x,y)=P_2(x,y)$ for some
$y\in\{0,1\}^n$, by querying $P$ on $(x,y)\in\{0,1\}^n\times\{0,1\}^n$.
\end{definition}

Notice that the 0-1 Knapsack problem is an instance of the symmetric
claw problem.

\begin{remark}
  Consider a simpler version of the 0-1 Knapsack problem, where we
  seek a 0-1 assignment to variables $x_1,\ldots,x_n$ such that
  $\sum_{i=1}^n a_ix_i=0$, given integers $a_1,\ldots,a_n$ as input.
  
  For each $y\in\{0,1\}^n$ define a partition of the variable set
  $A=\{i\in[n]\mid y_i=1\}$ and $B=\{i\in[n]\mid y_i=0\}$.  Define
  $P_1(x,y)=\sum_{i\in A} a_ix_i$ and $P_2(x,y)=-\sum_{i\in B} a_ix_i$.
  
  It is easy to check that $P_1$ and $P_2$ fulfill conditions of
  Definition~\ref{claw1def}, making this version of 0-1 Knapsack an instance
  of the symmetric claw problem.
\end{remark}

We first establish the upper bound result for symmetric claw problem.
It is the same algorithm we described in the previous section.

\begin{theorem}\label{ubound}
There is an $\O(2^{n/3})$ quantum algorithm with constant error
probability for the symmetric claw problem in the quantum query model.
\end{theorem}

\begin{proof}
  The algorithm is along exactly the same lines as the algorithm in
  Theorem~\ref{theorem1} for 0-1 ILP. In fact it is an easier version
  of that algorithm as we do not require range trees here.
  
  We fix $y\in\{0,1\}^n$ to be such that $y_i=1$ for $1\leq i\leq n/3$
  and $y_i=0$ for $n/3+1\leq i\leq n$.

Let $X_A=\{x\in\{0,1\}^n\mid x_i=0$ for $n/3+1\leq i\leq n\}$ and
$X_B=\{x\in\{0,1\}^n\mid x_i=0$ for $1\leq i\leq n/3\}$.

Compute and sort the list $\{P_1(x,y)\mid x\in X_A\}$ of size
$2^{n/3}$ using a classical sorting algorithm in time $\O(2^{n/3})$.
This is the preprocessing phase. 

Prepare a superposition of $x\in X_B$ and search for $P_2(x,y)$ in the
sorted list using a classical binary search implemented by a unitary
transform $U$ defined as: For $x\in X_B$, $U\ket{x}=1$ if $P_2(x,y)$
occurs in the sorted list and $U\ket{x}=0$ otherwise.

Now, using Grover's search we can find $x\in X_B$ such that
$P_1(x',y)=P_2(x,y)$ for some $x'\in X_A$. At this point, using the
properties of $P_1$ and $P_2$ we can put together a complete 0-1
assignment $x''$ from $x$ and $x'$ such that $P_1(x'',y)=P_2(x'',y)$.
We can obtain $x''$ by concatenating the first $n/3$ bits from $x'$
and the last $2n/3$ bits from $x$.
\end{proof}

We now turn to the question of lower bounds. Using the technique of
Ambainis \cite{ambain} we establish a lower bound of $\Omega(2^{n/4})$
for the quantum query complexity of the symmetric claw problem. We
first state the result in a form that we need for our setting.

\begin{theorem}{\rm\cite{ambain}}\label{amb}
  Let ${\cal F}$ be the set of input functions $f$, where
  $f:[M]\rightarrow[M]$, and let $\phi:{\cal F}\rightarrow\Z\times\Z$
  be the function which we wish to compute. Let $X,Y$ be two subsets
  of ${\cal F}$ such that $\phi(f)\neq \phi(g)$ for all $f\in X$ and
  $g\in Y$.  Let $R\subseteq X\times Y$ such that:
\begin{enumerate}
\item[(i)] For every $f\in X$ there are at least $m$ different $g\in
  Y$ such that $(f,g)\in R$.
\item[(ii)] For every $g\in Y$ there are at least $m'$ different $f\in
  X$ such that $(f,g)\in R$.
\item[(iii)] For every $f\in X$ and $x\in[N]$, there are at most $l$
  different $g\in Y$ such that $(f,g)\in R$ such that $f(x)\neq g(x)$.
\item [(iv)] For every $g\in Y$ and $x\in[N]$, there are at most $l'$
  different $f\in X$ such that $(f,g)\in R$ such that $f(x)\neq g(x)$.
\end{enumerate}

Any quantum algorithm that evaluates $\phi$ with constant success
probability must make $\Omega(\sqrt{\frac{mm'}{ll'}})$ queries to the
input function in the quantum query model.
\end{theorem}

Now we are ready to prove our lower bound result of this section.

\begin{theorem}\label{lbound}
Any quantum algorithm that solves the symmetric claw problem with
constant success probability needs $\Omega(2^{n/4})$ queries in
the quantum query model.
\end{theorem}

\begin{proof}
  We will prove the claimed lower bound by applying
  Theorem~\ref{amb}. The inputs consist of the class ${\cal F}$ of
  functions
  $P:\{0,1\}^n\times\{0,1\}^n\rightarrow\{0,1\}^n\times\{0,1\}^n$,
  where we write $P(x,y)=(P_1(x,y),P_2(x,y))$, and every $P$ in ${\cal
    F}$ fulfills the conditions of the Definition~\ref{claw1def}. The
  goal of a quantum algorithm $\phi$ for the symmetric claw problem is
  to find $x\in\{0,1\}^n$ such that $P_1(x,y)=P_2(x,y)$ for some (and
  therefore, by the promise of Definition~\ref{claw1def}, every $y$).
  Let $\no$ denote the subset of ``no'' instances of ${\cal F}$ and
  $\yes$ the subset of ``yes'' instances. Define $X\subseteq\no$ as:
  
  $X=\{P\in{\cal F}\mid P_1(x,y)\neq P_2(x,y)~\forall x,y\in\{0,1\}^n,
  P_1(x,y)\neq P_1(x',y)$ if $x|_y\neq x'|_y$ and $P_2(x,y)\neq
  P_2(x',y)$ if $x|_{\overline{y}}\neq x'|_{\overline{y}}\}$.
  
  We will define the subset $Y$ of $\yes$ by first defining a relation
  $R\subseteq X\times\yes$ and then setting $Y=\{P'\in\yes\mid
  (P,P')\in R$ for some $P\in X\}$. We now define the relation $R$.
  
  Given $P\in X$, for every $x\in\{0,1\}^n$ we define a distinct
$P'\in\yes$ and include $(P,P')$ in $R$. Given $P\in X$ and
$x\in\{0,1\}^n$, $P'$ is obtained from $P$ as follows:

\begin{itemize}
\item If $\sum_{i=1}^n y_i\geq n/2$ then define $P'_1(x,y)=P_1(x,y)$,
  and $P'_2(x,y)=P_1(x,y)$. Furthermore, for every $x'\in\{0.1\}^n$
  such that $x'|_{\overline{y}}=x|_{\overline{y}}$ define
  $P'_2(x',y)=P_1(x,y)$. For such $x'$, notice that by definition of
  $X$ we have $P'_1(x',y)\neq P_1(x,y)$.
  
\item If $\sum_{i=1}^n y_i < n/2$ then define $P'_2(x,y)=P_2(x,y)$,
  and $P'_1(x,y)=P_2(x,y)$. Furthermore, for every $x'\in\{0.1\}^n$
  such that $x'|_y=x|_y$ define $P'_1(x',y)=P_2(x,y)$. For such $x'$,
  notice that by definition of $X$ we have $P'_2(x',y)\neq P_2(x,y)$.

\item For all other pairs $(u,v)\in\{0,1\}^n\times\{0,1\}^n$
  define $P'_1(u,v)=P_1(u,v)$ and $P'_2(u,v)=P_2(u,v)$.
\end{itemize}

Let $P'$ be defined as above for a given $P\in X$ and $x\in\{0,1\}^n$.
Notice that by construction $P'$ is in $\yes$ and, furthermore, $x$ is
the unique element of $\{0,1\}^n$ such that $P'_1(x,y)=P'_2(x,y)$ for
$y\in\{0,1\}^n$. Thus we define $2^n$ distinct elements $P'\in\yes$,
one for every $x\in\{0,1\}^n$, such that $(P,P')\in R$. Therefore, in
the terminology of Theorem~\ref{amb}, $m=2^n$. For the remainder of
the proof, we denote by $P'_x$ the element $P'\in\yes$ defined as
above from an $x\in\{0,1\}^n$ and a given $P\in X$.

We next show that for any $P\in X$ and
$(x,y)\in\{0,1\}^n\times\{0,1\}^n$ there are at most $2^{n/2}$
elements $P'\in Y$ such that $P(x,y)\neq P'(x,y)$ and $(P,P')\in R$.
In terms of Theorem~\ref{amb} this implies $l\leq 2^{n/2}$. We
argue this in two cases:
\begin{enumerate}
\item[Case 1.] Suppose $\sum_{i=1}^n y_i\geq n/2$. Then notice from
  the definition of $R$ that $P'_z$ differs from $P$ at $(x,y)$
  exactly when $x|_y=z|_y$. Thus, there are at most $2^{n/2}$ such
  elements $P'_z$.
\item[Case 2.] Suppose $\sum_{i=1}^n y_i < n/2$. Again, notice from
  definition of $R$ that $P'_z$ differs from $P$ at $(x,y)$ exactly
  when $x|_{\overline{y}}=z|_{\overline{y}}$. Thus, again, there can
  be at most $2^{n/2}$ such elements $P'_z$.
\end{enumerate}

Finally, notice that $l'\leq m'$ always holds. 

We can now apply the lower bound result Theorem~\ref{amb}. Since
\[
\frac{mm'}{ll'}\geq \frac{m}{l}\geq \frac{2^n}{2^{n/2}},
\]

The lower bound of $\Omega(\sqrt{\frac{mm'}{ll'}})$ given by
Theorem~\ref{amb} yields $\Omega(2^{n/4})$. This completes the proof.
\end{proof}

\section{Quantum Algorithm for CNF-SAT}

In this section we use ideas from Section~\ref{ilp} to give a simple
quantum search algorithm for CNF-SAT when the number of clauses is
linearly bounded in the number of variables. More precisely, for any
constant $c>0$ we define the following class of CNF formulas:
$\CNF_c=\{F\mid F$ is in conjunctive normal form with $n$ variables
and $F=C_1\wedge C_2\wedge\ldots\wedge C_m$ where $m\leq cn\}$.

\begin{theorem}\label{cnfsat}
  For any constant $c>0$ there is an $\O(2^{(1-\alpha)n/2})$ quantum
  algorithm for satisfiability for inputs $F\in\CNF_c$, where $\alpha$
  is a constant such that $\alpha<1/6$ and
  $H(\alpha)\leq\frac{1}{4c}$. Here, $H(\alpha)$ is the entropy
function defined as $-\alpha\log_e\alpha -(1-\alpha)\log_e(1-\alpha)$.
\end{theorem}

\begin{proof}
  Let $\alpha<1/6$ be a constant to be fixed later, as claimed in the
  statement. Let the input instance from $\CNF_c$ be $F=C_1\wedge
  C_2\wedge\ldots\wedge C_m$ from $\CNF_c$ with variables
  $x_1,x_2,\ldots,x_n$. Let $\frac{1}{\alpha}=k$, where we assume for
  simplicity of analysis that $k$ is an integer. Partition the
  variable set $\{x_1,x_2,\ldots,x_n\}$ into $k$ equal size sets
  $A_1,A_2,\ldots,A_k$, where $|A_i|=\alpha n$, $1\leq i\leq k$.
  Suppose $F$ is satisfiable and $a^*$ is some fixed satisfying
  assignment for $F$. Denote by $\ov{A_i}$ the set of variables
  $\{x_1,\ldots,x_n\}\setminus A_i$ for each $i$. Denote by $b^*_i$
  the partial assignment given by $a^*$ when restricted to variables in
  $A_i$. Similarly, let $c^*_i$ denote the partial assignment given by
  $a^*$ when restricted to variables in $\ov{A_i}$.\\

\begin{claim}
  There is an $i:~1\leq i\leq k$ such that the partial assignment
  $c^*_i$ satisfies at least $(1-\alpha)m$ clauses of $F$.
\end{claim}

Suppose the claim is false. Then $\ov{A_i}$ satisfies at most
$(1-\alpha)m$ clauses of $F$ for each $i$. Since $a^*$ is a satisfying
assignment, for each $i$ there is a set $S_i$ of \emph{more than}
$\alpha m$ clauses satisfied by the partial assignment $b^*_i$ such
that $S_i\cap S_j=\emptyset$ for $i\neq j$. This is impossible since
there are only $m$ clauses in $F$. The claim follows.

We now give a step-wise description of the algorithm. The algorithm is
a loop with index $i:~1\leq i\leq k$, where in the $i$th iteration it
considers the partition of the variable set into $A_i$ and $\ov{A_i}$.
The algorithm succeeds when it considers a partition $A_i$ and
$\ov{A_i}$ such that $c^*_i$ satisfies at least $(1-\alpha)m$ clauses
of the input $F$ (such an index exists by the claim).

\begin{enumerate}
\item[Step 1.] Partition the variable set into $A_i$ and $\ov{A_i}$.
  
\item[Step 2.] For $1\leq l\leq \alpha m$ build a sorted table $T_l$
  consisting of the following set of pairs $(u,b)$: $u\in\{0,1\}^m$ is
  am $m$-bit string with exactly $l$ 1's and $b$ is a truth assignment
  to variables in $A_i$ such that $b$ satisfies each clause $C_j$ for
  which $u_j=1$, where $1\leq j\leq m$. All other pairs $(u,b)$ are
  discarded. Then, using a classical sorting algorithm and treating
  $u$ in $(u,b)$ as an $m$-bit integer key, sort the pairs in $T_l$ in
  increasing order.  This entire step can be done in time $\O({m
    \choose \alpha n}2^{\alpha n})$.
  
\item[Step 3.] Using the Hadamard transform prepare a uniform
  superposition of the truth assignments $v$ to variables in $\ov{A_i}$.
  Let $S$ be the set of all such truth assignments. Notice that
  $|S|=2^{(1-\alpha)n}$. The superposition is
\[
\ket{\psi}=\sqrt{\frac{1}{2^{(1-\alpha)n}}}\sum_{v\in S}\ket{v}.
\]

\item[Step 4.] Define a unitary transform $U$ as the standard
  reversible implementation of the following classical subroutine $f$:
\begin{itemize}
\item[(i)] Compute the vector $u\in\{0,1\}^m$ where $u_j=1$ if and
  only if $C_j$ is not satisfied by the partial assignment $v$.
\item[(ii)] If $u$ has more than $\alpha m$ 1's then $f(v)$ returns
  value $0$ and stop.
  
\item[(iii)] Otherwise, if $u$ has $l$ 1's in it, $1\leq l\leq \alpha
  m$, then do a binary search for $u$ in the table $T_l$. If $(u,b)$
  is found in $T_l$ for some $b$ then $f(v)=1$ else $f(v)=0$. Notice
  that if $f(v)=1$ then $v$ and $b$ together give a satisfying
  assignment for $F$.
\end{itemize}

\item[Step 5.] Now, starting with initial state as $\ket{\psi}$ and
  using the unitary transform $U$ to implement $f$, apply Grover
  search to find $v\in S$ such that $f(v)=1$.
  
\item[Step 6.] Output the satisfying assignment $(b,v)$.

\end{enumerate}

To argue correctness it suffices to notice by our earlier claim that
for some $i$, $A_i$ and $\ov{A_i}$ is a partition such that in Step 3,
$f(v)=1$ for at least the partial assignment $v=c^*_i$ obtained from
$a^*$.

Thus, if $F$ is satisfiable the algorithm will find a satisfying
assignment with constant success probability in time $\O({m \choose
  \alpha m}2^{\alpha n}) + \O(2^{(1-\alpha)n/2})$, where the first
term is the preprocessing time in Step 2, and the second term is the
time for the Grover search. 

Since $m\leq cn$, an easy analysis yields ${m \choose \alpha m}\leq
2^{H(\alpha)cn}$ for large $n$. Thus, if we choose $\alpha<1/6$ such
that $(1-\alpha)/2\geq H(\alpha)c+\alpha$, then the overall time taken
is $\O(2^{(1-\alpha)n/2})$. It suffices to choose $\alpha$ such that
$H(\alpha)<\frac{1}{4c}$.
\end{proof}

\section{Discussion}\label{claw2}

The question that arises is whether we can close the gap between the
upper and lower bounds for the symmetric claw problem. Also, for the
0-1 ILP problem with $n$ variables and $d$ inequalities we do not have
a better upper bound than $\O(2^{n/3})n^d$. To study this issue in the
quantum query model we define the simultaneous claw problem.
 
\begin{definition}\label{simclaw}
Let $(f_i,g_i), 1\leq i\leq d$ be functions where $f_i:[N]\rightarrow
X$ and $g_i:[N]\rightarrow X$, where $X$ is some set. The
\emph{simultaneous claw} problem is to find an $x\in[N]$ such that
$f_i(x)=g_i(x)$ for each $i=1,2,\ldots,d$.
\end{definition}

We can also define a symmetric version of the simultaneous claw
problem similar to Definition~\ref{claw1def}, but for simplicity we
focus on this definition. It is known \cite{buhretal} that for
$d=1$ there is an $\O(N^{3/4})$ upper bound and the recent new
techniques of Aaronson \cite{aaron} followed by Shi's sharpened
results \cite{shi} imply an $\Omega(N^{2/3})$ lower bound for the
problem.

We show here that the simultaneous claw problem has a quantum upper
bound of $\O(N^{3/4}\log^d N)$. On the other hand, we are not able to
strengthen the lower bound of $\Omega(N^{2/3})$ which already holds
for a single claw pair $(f,g)$. The following theorem is easy to prove.

\begin{theorem}
There is a quantum algorithm that takes as input a collection
$(f_i,g_i), 1\leq i\leq d$ of functions where $f_i:[N]\rightarrow X$
and $g_i:[N]\rightarrow X$, where $X$ is some set, which makes
$\O(N^{3/4}\log^d N)$ queries in the quantum query model and outputs a
simultaneous claw $x\in[N]$, if it exists, such that $f_i(x)=g_i(x)$
for each $i=1,2,\ldots,d$.
\end{theorem}

We recall the $r$-to-1 collision problem: given as input a function
$f:[N]\rightarrow X$ with the promise that $f$ is either 1-1 or
$r$-to-1, the problem is to find a pair $x\neq y\in[N]$ such that
$f(x)=f(y)$. The recent results of Aaronson \cite{aaron} followed by
Shi \cite{shi} imply an $\Omega(N^{1/3})$ lower bound for the 2-to-1
collision problem.

Analogous to the simultaneous claw problem we can define the
simultaneous 2-to-1 collision problem which is given as input
functions $f_i:[N]\rightarrow X$, $1\leq i\leq d$, with the promise
that either all the $f_i$ are 1-1 or all are 2-to-1, find a pair
$x\neq y\in[N]$ if it exists such that $f_i(x)=f_i(y)$ for each $i$.
The $\Omega(N^{1/3})$ lower bound of Shi \cite{shi} clearly holds but
we have nothing better than that. On the other hand, we have a
straightforward algorithm that gives an upper bound for the problem.

\begin{theorem}
There is a quantum algorithm that takes as input a collection $f_i,
1\leq i\leq d$ of functions $f_i:[N]\rightarrow X$ where $X$ is some
set, makes $\O(N^{3/4}\log^d N)$ queries in the quantum query model
and outputs a simultaneous collision $x\neq y\in[N]$ such that
$f_i(x)=f_i(y)$ for each $i=1,2,\ldots,d$, if it exists.
\end{theorem}


\end{document}